\documentclass[aps,prb,twocolumn,showpacs,floatfix]{revtex4}
\usepackage{amsmath}
\usepackage{amssymb}
\usepackage{bm}
\usepackage{graphicx}

\begin{document}


\title
{Thermal broadening of the J-band in disordered linear molecular
aggregates:\\ A theoretical study}

\author{D.\ J.\ Heijs}

\author{V.\ A.\ Malyshev}
\thanks{On leave from S.I. Vavilov State Optical Institute,
Birzhevaya Liniya 12, 199034 Saint-Petersburg, Russia.}

\author{J.\ Knoester}
\affiliation{Institute for Theoretical Physics and  Materials
Science Center, University of Groningen, Nijenborgh 4, 9747 AG
Groningen, The Netherlands}

\date{\today}

\begin{abstract}

We theoretically study the temperature dependence of the J-band
width in disordered linear molecular aggregates, caused by
dephasing of the exciton states due to scattering on vibrations of
the host matrix. In particular, we consider inelastic one- and
two-phonon scattering between different exciton states
(energy-relaxation-induced dephasing), as well as elastic
two-phonon scattering of the excitons (pure dephasing). The
exciton states follow from numerical diagonalization of a Frenkel
Hamiltonian with diagonal disorder; the scattering rates between
them are obtained using the Fermi Golden Rule. A Debye-like model
for the one- and two-phonon spectral densities is used in the
calculations. We find that, owing to the disorder, the dephasing
rates of the individual exciton states are distributed over a wide
range of values. We also demonstrate that the dominant channel of
two-phonon scattering is not the elastic one, as is often tacitly
assumed, but rather comes from a similar two-phonon inelastic
scattering process. In order to study the temperature dependence
of the J-band width, we simulate the absorption spectrum,
accounting for the dephasing induced broadening of the exciton
states. We find a power-law ($T^p$) temperature scaling of the
effective homogeneous width, with an exponent $p$ that depends on
the shape of the spectral density of host vibrations. In
particular, for a Debye model of vibrations, we find $p\approx 4$,
which is in good agreement with experimental data on J-aggregates
of pseudoisocyanine [J. Phys. Chem. A {\bf 101}, 7977 (1997)].

\end{abstract}


\maketitle

\section{Introduction}
\label{introduction}

Ever since the discovery of aggregation of cyanine dye molecules
by Jelley~\cite{Jelley36} and Scheibe~\cite{Scheibe36} the width
of the absorption band of these linear aggregates (the J-band) has
attracted much attention. At low temperatures, the J-band may be
as narrow as a few tens of cm$^{-1}$'s (for pseudoisocyanine),
while at room temperature it typically is a few hundred
cm$^{-1}$.\cite{Kobayashi96} The small width at low temperature is
generally understood as resulting from the excitonic nature of the
optical excitations,\cite{Franck38} which leads to exchange
narrowing of the inhomogeneous broadening of the transitions of
individual molecules.\cite{Knapp84} As the optically dominant
exciton states in inhomogeneous J-aggregates occur below the
exciton band edge, their vibration-induced dephasing is strongly
suppressed at low temperatures. Thus, the J-band is
inhomogeneously broadened, except for a small residual homogeneous
component ($\sim$ 0.1 cm$^{-1}$) caused by spontaneous emission of
the individual exciton states underlying the spectrum. Upon
increasing the temperature, the vibration-induced dephasing of the
exciton states increases, the J-band broadens, and obtains a more
homogeneous character.

Over the past twenty years, the temperature dependence of the
J-band width and the homogeneous broadening of the exciton states,
has been studied by several authors. In 1987, De Boer, Vink, and
Wiersma~\cite{deBoer87} performed accumulated-echo experiments to
study the temperature dependence of the pure dephasing time in
J-aggregates of pseudo-isocyanine bromide (PIC-Br) and found that
within the temperature range 1.5 K to 100 K, this time is linearly
proportional to the occupation number of a mode with an energy of
9 cm$^{-1}$. They assigned this to a librational mode of the
aggregate. Later on, Fidder, Knoester, and Wiersma~\cite{Fidder90}
(see also Ref.~\onlinecite{Fidder91} and~\onlinecite{Fidder93})
showed that a similar study carried our over a wider temperature
range, 1.5 K to 190 K, required the occupation numbers of three
vibration modes, at 9 cm$^{-1}$, 305 cm$^{-1}$, and 973 cm$^{-1}$.
Using the hole burning technique, Hirschmann and
Friedrich~\cite{Hirschmann89} studied the homogeneous width of the
exciton states in pseudoisocyanine iodide (PIC-I) over the
temperature range 350 mK to 80 K. They were able to fit their
measurements by a superposition of two exponentials, with
activation energies 27 cm$^{-1}$ and 330 cm$^{-1}$, and they
attributed the broadening to scattering of the excitons on an
acoustic mode and an optical mode, respectively, of the aggregate.
Finally, in 1997 Renge and Wild~\cite{Renge97} measured the
temperature dependent width, $\Delta(T)$, of the total J-band of
pseudoisocyanine chloride (PIC-Cl) and fluoride (PIC-F) over the
wide temperature range 10 K to 300 K. They found that over this
entire range their data closely obeyed a power-law dependence
$\Delta(T) = \Delta(0) + bT^p$ with the exponent $p = 3.4$. They
suggested the scattering of the excitons on host vibrations as a
possible source of this behavior.

The above overview clearly demonstrates that the temperature
dependence of the most important characteristic of J-aggregates,
namely the J-band, is not understood. It even is not clear what is
the source of thermal broadening: dephasing due to vibrational
modes of the aggregate itself or due to modes of the host matrix.
Theoretically, the study of the exciton dephasing in J-aggregates
is complicated by the fact that static disorder plays an important
role in these systems, as is clear from the strongly asymmetric
low-temperature lineshape.~\cite{Fidder91,Fidder93} The
simultaneous treatment of disorder, leading to exciton
localization, and scattering on vibrational modes, is a problem
that requires extensive numerical simulations. Such simulations
have been used previously to model the optical response of
polysilanes,~\cite{Shimitsu98,Shimizu01} the temperature dependent
fluorescence~\cite{Bednarz03} and transport~\cite{Malyshev03}
properties of J-aggregates, and single molecular spectroscopy of
circular aggregates.~\cite{Dempster01} Alternatively, stochastic
models have been applied to describe the scattering of excitons on
vibrations in disordered aggregates and the resulting spectral
line shape~\cite{Barvik99,Warns03} and relaxation
dynamics.~\cite{Lemaistre99}

In this paper, we report on a systematic theoretical study of
dephasing of weakly localized Frenkel excitons in one-dimensional
systems, focusing on the effect of the scattering of the excitons
on the vibrational modes (phonons) of the host. For a chain-like
configuration it seems physically reasonable to assume that the
coupling to host vibrations dominates the dynamics of the
excitons. For coupling to vibrations in the chain, one expects
self-trapped exciton states,~\cite{Agranovich99} for which in most
aggregates no clear signature is found. We describe the phonons by
a Debye model and consider one-phonon as well as (elastic and
inelastic) two-phonon contributions to the dephasing. The
scattering rates between the various, numerically obtained,
exciton states are derived using the Fermi Golden Rule. Due to the
disorder, the dephasing rates of individual exciton states are
distributed over a wide range, in particular at low temperature,
making it meaningless to associate the width of individual exciton
levels with the J-band width. Rather, this width is determined by
direct simulation of the total absorption spectrum and is found to
scale with temperature according to a power law.

The outline of this paper is as follows: In Sec.~\ref{model} we
introduce the Hamiltonian for excitons in a disordered chain and
coupled to host vibrations. General expressions for the exciton
dephasing rates are derived in Sec.~\ref{dephasing}.
Section~\ref{homogeneouslimit} deals with the temperature
dependence of the dephasing rates for the disorder-free case,
where analytical expressions can be obtained. In
Sec.~\ref{numerics} we give the results of our numerical
simulations for the dephasing rates in the presence of disorder
and analyze their temperature dependence and fluctuations as well
as the total J-band width. We compare to experiment in
Sec.~\ref{experiment} and discuss an alternative mechanism of
dephasing due to the coupling of excitons to a local vibration. In
Section~\ref{concl} we present our conclusions.

\section{Model}
\label{model}

We consider an ensemble of J-aggregates embedded in a disordered
host matrix. The aggregates are assumed to be decoupled from each
other, while they interact with the host. A single aggregate is
modeled as an open linear chain of $N$ coupled two-level monomers
with parallel transition dipoles. The interaction between a
particular monomer and the surrounding host molecules in the
equilibrium configuration leads to shifts in the monomer's
transition energy. Due to the host's structural disorder, this
shift is different for each monomer in the aggregate, giving rise
to on-site (diagonal) disorder. Moreover, vibrations of the host
couple to the aggregate excited states, because the associated
displacements away from the equilibrium configuration dynamically
affect the monomer transition energies. Accounting for these
shifts up to second order in the molecular displacements, the
resulting Hamiltonian in the site representation reads:
\begin{subequations}
\label{H}
\begin{equation}
    H = H^\mathrm{ex} + H^\mathrm{bath} + V^{(1)} + V^{(2)} \ ,
\end{equation}
with
\begin{equation}
\label{Hex}
    H^\mathrm{ex}
    = \sum_{n=1}^N \> \varepsilon_n |n\rangle \langle n|
    + \sum_{n,m=1}^N\> J_{nm} \> |n\rangle \langle m| \ ,
\end{equation}
\begin{equation}
\label{Hbath}
    H^\mathrm{bath}
    = \sum_q \> \omega_q a^{\dag}_q a_q \ ,
\end{equation}
\begin{equation}
\label{V1}
    V^{(1)} = \sum_{n=1}^N \> \sum_q V^{(1)}_{nq}
    |n\rangle \langle n|(a_q + a_q^{\dag})  \ ,
\end{equation}
\begin{equation}
\label{V2}
    V^{(2)} = \sum_{n=1}^N \> \sum_{qq^{\prime}}
    V^{(2)}_{nqq^{\prime}} |n\rangle \langle n|
    (a_q + a_q^{\dag})
    (a_{q^{\prime}} + a_{q^{\prime}}^{\dag}).
\end{equation}
\end{subequations}
Here, $H^\mathrm{ex}$ is the bare Frenkel exciton Hamiltonian,
with $|n \rangle$ denoting the state in which the $n$th monomer is
excited and all the other monomers are in the ground state. The
monomer excitation energies, $\varepsilon_1,\varepsilon_2, \ldots
, \varepsilon_N$, are uncorrelated stochastic gaussian variables,
with mean $\bar\varepsilon$ and standard deviation $\sigma$,
referred to as the disorder strength. Hereafter, $\bar\varepsilon$
is set to zero. The resonant interactions $J_{nm}$ are considered
to be nonrandom and are assumed to be of dipolar origin: $J_{nm} =
- J/|n-m|^{3}$ \, $(J_{nn} \equiv 0)$, with $J > 0$ denoting the
nearest-neighbor coupling.

$H^\mathrm{bath}$ describes the vibrational modes of the host,
labeled $q$ and with the spectrum $\omega_q$ ($\hbar = 1$). The
operator $a_q \, (a^{\dag}_q)$ annihilates (creates) a vibrational
quantum in mode $q$. Finally, the operators $V^{(1)}$ and $V^{(2)}$
describe the linear and quadratic exciton-vibration coupling,
respectively, where the quantities $V^{(1)}_{nq}$ and
$V^{(2)}_{nqq^{\prime}}$ indicate their strengths. We do not derive
explicit expressions for these coupling strengths, as we aim to
treat them on a phenomenological basis. In particular, owing to
the disordered nature of the host, we consider these strengths
stochastic quantities, for which we only specify the following
stochastic properties with respect to the site index $n$:
\begin{subequations}
\label{VV}
\begin{equation}
\label{<V>}
    \left\langle V^{(1)}_{nq} \right\rangle
    = \left\langle V^{(2)}_{nqq^{\prime}} \right\rangle = 0 \ ,
\end{equation}
\begin{equation}
\label{<V1V1>}
    \left\langle V^{(1)}_{mq}V^{(1)*}_{nq} \right\rangle
    = \delta_{mn} {\left|V^{(1)}_q\right|}^2  \ ,
\end{equation}
\begin{equation}
\label{<V2V2>}
    \left\langle V^{(2)}_{mqq^{\prime}}
    V^{(2)*}_{nqq^{\prime}} \right\rangle
    = \delta_{mn} {\left|V^{(2)}_{qq^{\prime}}\right|}^2  \ ,
\end{equation}
\end{subequations}
where the angular brackets denote averaging over realizations of
$V^{(1)}_{nq}$ and $V^{(2)}_{nqq^{\prime}}$. The Kronecker symbol
in Eqs.~(\ref{<V1V1>}) and~(\ref{<V2V2>}) implies that the
surroundings of different monomers in the aggregate are not
correlated.

In general, the Hamiltonian Eq.~(\ref{H}) can not be diagonalized
analytically. If the exciton-vibration coupling is not too strong,
the method of choice is first to find the exciton eigenstates by
numerical diagonalization and then to consider the scattering on
the basis of these eigenstates. Explicitly, the exciton states
follow from the eigenvalue problem
\begin{subequations}
\label{diagonalization}
\begin{equation}
    \sum_{m=1}^N H^\mathrm{ex}_{nm} \varphi_{\nu m}
    = E_\nu \varphi_{\nu n} \ , \quad \nu = 1,2, \ldots , N \ ,
\label{eigenen}
\end{equation}
where $H^\mathrm{ex}_{nm} = \langle n| H^\mathrm{ex} |m\rangle$
and $E_{\nu}$ is the eigenenergy of the exciton state $|\nu
\rangle$:
\begin{equation}
    |\nu\rangle=\sum_{n=1}^N \varphi_{\nu n}|n\rangle \ .
\label{eigenfu}
\end{equation}
\end{subequations}
In the exciton representation, the exciton-vibration interactions
take the form
\begin{subequations}
\label{V1V2}
\begin{equation}
\label{a}
    V^{(1)} = \sum_{\mu,\nu=1}^N \> \sum_q V^{(1)}_{\mu\nu q}
    |\mu \rangle \langle \nu| (a_q + a_q^{\dag})  \ ,
\end{equation}
\begin{equation}
\label{b}
    V^{(2)} = \sum_{\mu,\nu=1}^N \> \sum_{qq^{\prime}}
    V^{(2)}_{\mu\nu qq^{\prime}} |\mu\rangle \langle \nu|
    (a_q + a_q^{\dag})(a_{q^{\prime}} + a^{\dag}_{q^{\prime}}) \ ,
\end{equation}
\end{subequations}
where $V^{(1)}_{\mu\nu q}$ and  $V^{(2)}_{\mu\nu qq^{\prime}}$ are
the matrix elements of the vibration-induced scattering of an
exciton from state $|\nu \rangle$ to state $|\mu \rangle$, given by
\begin{subequations}
\label{V1qV2q}
\begin{equation}
\label{V1q}
    V^{(1)}_{\mu\nu q} = \sum_{n=1}^N \> V^{(1)}_{nq}
    \varphi_{\mu n} \varphi_{\nu n} \ ,
\end{equation}
\begin{equation}
\label{V2q}
    V^{(2)}_{\mu\nu qq^{\prime}} = \sum_{n=1}^N \>
    V^{(2)}_{nqq^{\prime}} \varphi_{\mu n} \varphi_{\nu n} \ .
\end{equation}
\end{subequations}

Spectroscopic data on J-aggregates clearly reveal that the
exciton-vibration coupling in these systems is usually weak. For
the prototypical J-aggregates of PIC, this claim is corroborated
by two facts: (i) - the narrowness of the J-band, which only is a
few tens of cm$^{-1}$ at liquid helium temperature and becomes
several times broader at room temperature, and (ii) - the absence
of a fluorescence Stokes shift of the J-band (see, e.g.,
Ref.~\onlinecite{Fidder90}). The extended nature of the exciton
states in J-aggregates helps to reduce the exciton-vibration
coupling, as it leads to averaging of the static as well as
dynamic fluctuations of the site energies, effects known as
exchange~\cite{Knapp84} and motional~\cite{Wubs98,Malyshev98}
narrowing, respectively. The weakness of the exciton-vibration
coupling allows one to calculate the scattering and dephasing
rates of the excitons through perturbation theory. This analysis
is presented in the next section.

\section{Dephasing rates}
\label{dephasing}

Following the arguments given at the end of Sec.~\ref{model}, we
will use Fermi's Golden Rule to calculate the rate for scattering
of excitons from one localized state, $|\nu \rangle$, to another
one, $|\mu \rangle$. The result reads
\begin{eqnarray}
\label{FGR}
    W_{\mu\nu}^{(\xi)}
    & = &2\pi \sum_f \sum_i \rho(\Omega_i)
    \left \langle \left| \langle \mu, f|V^{(\xi)}|\nu,i \rangle
    \right|^2 \right\rangle
\nonumber\\
\nonumber\\
    & \times & \delta(E_\mu - E_\nu
    + \Omega_f - \Omega_i) \ .
\end{eqnarray}
Here, the superscript $\xi = 1,2$ distinguishes between one- and
two-vibration-assisted exciton scattering. Furthermore, $\Omega_i$
and $\Omega_f$ are the energies of the vibration bath in the
initial ($|i \rangle = | \{n_q\}_i \rangle$) and final ($|f
\rangle = | \{n_q\}_f \rangle$) states, respectively, where
$\{n_q\}$ denotes the set of occupation numbers of the vibrational
modes. The quantity $\rho(\Omega_i)$ is the equilibrium density
matrix of the initial state of the bath. Finally, the angular
brackets indicate that we average over the stochastic realizations
of the surroundings of each monomer in the aggregate.

\subsection{Linear exciton-vibration coupling}
\label{linear}

In a one-phonon-assisted scattering process, the occupation number
of one phonon mode $q$ increases or decreases by one, corresponding
to emission and absorption of a vibrational quantum, respectively.
Consequently, $\Omega_f - \Omega_i = \pm \omega_q$. Substituting
the explicit form of the operator $V^{(1)}$ from Eq.~(\ref{V1})
into Eq.~(\ref{FGR}), we obtain
\begin{eqnarray}
\label{W1}
    W^{(1)}_{\mu\nu}
    & = & 2\pi \sum_{n=1}^N \varphi^2_{\mu n} \varphi_{\nu n}^2
    \sum_q \left|V^{(1)}_q \right|^2 \,
\nonumber\\
\nonumber\\
    & \times & \Big[
    \big[\bar{n}(\omega_q) + 1\big]\, \delta(\omega_{\mu\nu} + \omega_q)
\nonumber\\
\nonumber\\
    & + & \bar{n}(\omega_q) \, \delta(\omega_{\mu\nu} - \omega_q)\Big] \ ,
\end{eqnarray}
where $\bar{n}(\omega_q) = [\exp(\omega_q/T) - 1]^{-1}$ is the
mean occupation number of the vibrational mode $q$ (the Boltzmann
constant $k_B = 1$) and $\omega_{\mu\nu} = E_\mu - E_\nu$. In
deriving Eq.~(\ref{W1}), we used the properties of the stochastic
function $V^{(1)}_{nq}$ given by Eqs.~(\ref{<V>}) and
(\ref{<V1V1>}). Defining the one-vibration spectral density as
\begin{equation}
\label{F1}
    {\cal F}^{(1)}(\omega) \equiv 2\pi \sum_q \left|V^{(1)}_q \right|^2
    \delta(\omega - \omega_q) \ ,
\end{equation}
we can rewrite Eq.~(\ref{W1}) in the form
\begin{eqnarray}
\label{w1a}
    W^{(1)}_{\mu\nu}
    & = &  \sum_{n=1}^N \varphi_{\mu n }^2 \varphi_{\nu n}^2 \,
    {\cal F}^{(1)}(|\omega_{\mu\nu}|)
\nonumber\\
\nonumber\\
& \times & \left\{
\begin{array}{lr}
\bar{n}(\omega_{\mu\nu}), &\omega_{\mu\nu} > 0 \ ,
\\
\\
\bar{n}(-\omega_{\mu\nu}) +1 , &\omega_{\mu\nu} < 0 \ .
\end{array}
\right.
\end{eqnarray}

As we observe, the rate $W^{(1)}_{\mu\nu}$ is proportional to the
overlap integral of the site occupation probabilities,
$\varphi^2_{\mu n}$ and $\varphi^2_{\nu n}$, of the exciton states
involved.  First of all, this leads to a strong suppression of the
scattering rate if states $|\mu \rangle$ and $|\nu \rangle$ overlap
weakly or not at all. Second, as the low-energy exciton states in a
disordered chain exhibit large fluctuations in their localization
size,\cite{Malyshev01} also the scattering rates may undergo large
fluctuations (see Sec.~\ref{numerics}).

The dependence of $W^{(1)}_{\mu\nu}$ on the energy mismatch
$\omega_{\mu \nu}$ is determined by the one-phonon spectral density
${\cal F}^{(1)}(\omega)$. Characterizing this function requires
knowledge of the vibrational spectrum $\omega_q$  as well as the
$q$ dependence of the exciton-vibration coupling $V^{(1)}_q$. For
the special case of scattering on acoustic phonons of a relatively
long wavelength, we have ${\cal F}^{(1)}(\omega) \sim \omega^3$.
This behavior results from the $\omega^2$-dependence of the
density of states of acoustic phonons, combined with the fact that
in the long-wavelength limit $|V^{(1)}_q|^2 \sim
\omega_q$.\cite{Davydov71,Bednarz02} One may consider this a
Debye-like model, in which one replaces the summation over the
mode index $q$ by an integration over the frequency $\omega_q$
according to the well-known rule:
\begin{equation}
\label{sum-to-int}
    \sum_q \to C \int_0^{\omega_c} d\omega_q \omega_q^2 \ .
\end{equation}
Here, $C$ is an irrelevant constant, which we will incorporate in
an overall free parameter (see below), and $\omega_c$ is a cutoff
frequency. It is important to note that $\omega_c$ is not
necessarily related to the Debye frequency: the generally very
complex density of vibrational states in a disordered solid may on
average exhibit an $\omega^2$ scaling up to a given frequency
$\omega_c$.

Inspired by the above, we consider a slightly wider class of
one-phonon spectral densities, given by
\begin{equation}
\label{F1model}
    {\cal F}^{(1)}(\omega)
    = W_0^{(1)} \left(\frac{\omega}{J}\right)^\alpha \,
    \Theta(\omega_c - \omega) \ .
\end{equation}
Here, $W_0^{(1)}$ is a free parameter in the model, which
characterizes the overall strength of the one-vibration-assisted
scattering rates. It absorbs a number of other parameters
characteristic for the host lattice (such as the velocity of
sound), the constant $C$ from Eq.~(\ref{sum-to-int}), as well as
the strength of the transfer interaction $J$ (for details, see
Ref.~\onlinecite{Bednarz02}). $\Theta(x)$ is the Heaviside step
function. When performing numerical simulations, we will mostly
use $\alpha = 3$, for which the spectral density of acoustic
phonons in the long-wave limit is recovered. In some instances,
however, we will discuss how results depend on the exponent
$\alpha$.

Several observations support considering a Debye-like vibration
spectral density, even for a disordered host. Thus, for strongly
disordered Yb$^{3+}$ doped phosphate glasses, a parabolic behavior
of the one-phonon spectral density was found over a rather wide
range of measurement (0 to 100 cm$^{-1}$).~\cite{Basiev87}
Furthermore, closely related spectral densities of the form ${\cal
F}^{(1)}(\omega) \sim (\omega/\omega_c)^{\alpha}
\exp(-\omega/\omega_c)$ (or linear combinations of such functions)
have been used successfully to fit the optical dynamics in
photosynthetic antenna complexes (see, e.g.,
Refs.~\onlinecite{Kuhn97,May00,Renger01,Brueggemann04}).

One-phonon-assisted scattering results in the transition of an
exciton from a given state $|\nu \rangle$ to state $|\mu \rangle$,
where necessarily $\mu \ne \nu$. In other words, this type of
scattering changes the occupation probabilities of the exciton
states and thus causes population (or energy) relaxation. The
population relaxation in turn contributes to the dephasing of state
$|\nu \rangle$. The corresponding dephasing rate is given by (see,
e.g., Ref.~\onlinecite{Blum96}):
\begin{equation}
\label{Gamma1}
    \Gamma^{(1)}_\nu \equiv \frac{1}{2} \, \sum_{\mu(\ne \nu)}
    W^{(1)}_{\mu\nu} \ .
\end{equation}
Thus, $\Gamma^{(1)}_\nu$ represents the one-phonon-assisted
contribution to the homogeneous broadening of the excitonic level
$\nu$. We note that the $\Gamma^{(1)}_\nu$ indirectly depend on
temperature through the $\bar{n}(\omega_{\mu\nu})$
[cf.~Eq.~(\ref{w1a})]. The temperature dependence of the sum over
scattering rates in Eq.~(\ref{Gamma1}) and the corresponding width
of the total exciton absorption spectrum will be analyzed in
Secs.~\ref{homogeneouslimit} and \ref{numerics}.

\subsection{Quadratic exciton-phonon coupling}
\label{quadratic}

When excitons scatter on the second-order displacements of the host
molecules, described by the operator $V^{(2)}$, the occupation
numbers of two phonon modes $q$ and $q^{\prime}$ with frequencies
$\omega_q$ and $\omega_{q^{\prime}}$ change by $\pm 1$. Thus,
$\Omega_f - \Omega_i = \pm \omega_q \pm \omega_{q^{\prime}}$,
where any combination of plus and minus is allowed. The
corresponding scattering rates $W^{(2)}_{\mu\nu}$ are obtained from
Eq.~(\ref{FGR}), taking into account the stochastic properties of
$V^{(2)}_{nqq^{\prime}}$ given by Eqs.~(\ref{<V>})
and~(\ref{<V2V2>}):
\begin{eqnarray}
\label{W2}
    W^{(2)}_{\mu\nu}
    & = & 2\pi \sum_{n=1}^N \varphi^2_{\mu n} \varphi_{\nu n}^2
    \sum_{qq^{\prime}} \left|V^{(2)}_{qq^{\prime}} \right|^2 \,
\nonumber\\
\nonumber\\
    & \times & \Big[\big[\bar{n}(\omega_q) + 1\big]
    \big[\bar{n}(\omega_{q^{\prime}}) + 1\big] \,
    \delta(\omega_{\mu\nu} + \omega_q + \omega_{q^{\prime}})
\nonumber\\
\nonumber\\
    & + & 2 \bar{n}(\omega_q)
    \big[\bar{n}(\omega_{q^{\prime}}) + 1\big] \,
    \delta(\omega_{\mu\nu} - \omega_q + \omega_{q^{\prime}})
\nonumber\\
\nonumber\\
    & + & \bar{n}(\omega_q) \bar{n}(\omega_{q^{\prime}}) \,
    \delta(\omega_{\mu\nu} - \omega_q - \omega_{q^{\prime}})\Big] \ .
\end{eqnarray}
If in analogy to the one-vibration-assisted scattering, we define
the two-vibration spectral density ${\cal
F}^{(2)}(\omega,\omega^{\prime})$ as
\begin{equation}
\label{F2}
    {\cal F}^{(2)}(\omega,\omega^{\prime}) \equiv 2\pi
    \sum_{qq^{\prime}} \left|V^{(2)}_{qq^{\prime}} \right|^2
    \delta(\omega - \omega_q) \delta(\omega^{\prime}
    - \omega_{q^{\prime}}) \ ,
\end{equation}
the scattering rate $W^{(2)}_{\mu\nu}$ takes the form
\begin{eqnarray}
\label{1W2}
    W^{(2)}_{\mu\nu}
    & = & \sum_{n=1}^N \varphi^2_{\mu n} \varphi_{\nu n}^2
    \int \mathrm{d}\omega \mathrm{d}\omega^{\prime}
    {\cal F}^{(2)}(\omega,\omega^{\prime})
\nonumber\\
\nonumber\\
    & \times & \Big[\big[\bar{n}(\omega) + 1\big]
    \big[\bar{n}(\omega^{\prime}) + 1\big] \,
    \delta(\omega_{\mu\nu} + \omega + \omega^{\prime})
\nonumber\\
\nonumber\\
    & + & 2 \bar{n}(\omega)
    \big[\bar{n}(\omega^{\prime}) + 1\big] \,
    \delta(\omega_{\mu\nu} - \omega + \omega^{\prime})
\nonumber\\
\nonumber\\
    & + & \bar{n}(\omega) \bar{n}(\omega^{\prime}) \,
    \delta(\omega_{\mu\nu} - \omega - \omega^{\prime})\Big] \ .
\end{eqnarray}
Similar to ${\cal F}^{(1)}(\omega)$ [Eq.~(\ref{F1})], we will use a
parametrization
\begin{equation}
\label{F2model}
    {\cal F}^{(2)}(\omega,\omega^{\prime})
    = \frac{W_0^{(2)}}{J}
    \left(\frac{\omega\omega^{\prime}}{J^2}\right)^\alpha \,
    \Theta(\omega_c - \omega)
    \Theta(\omega_c - \omega^{\prime})
     \ ,
\end{equation}
where $W_0^{(2)}$ is a free parameter that characterizes the
overall strength of the two-vibration-assisted scattering rates
[cf. $W_0^{(1)}$ and the discussion following Eq.~(\ref{F1model})].

Two main types of two-phonon-assisted processes may be
distinguished. Similarly to the one-phonon case, an inelastic
channel exists, where scattering occurs between different exciton
states, thus giving rise to population relaxation. However, also
an elastic channel is present, in which an exciton is scattered by
emitting and absorbing a phonon of the same energy, and the final
exciton state is identical to the initial one. This process
results in pure dephasing of the exciton state, with a rate given
by
\begin{eqnarray}
\label{Wnunu}
    W^{(2)}_{\nu\nu}
    = 2 \sum_{n=1}^N \varphi^4_{\nu n}
    \int \mathrm{d}\omega
    {\cal F}^{(2)}(\omega,\omega)
    \bar{n}(\omega) \big[\bar{n}(\omega) + 1\big] \ .
\end{eqnarray}
The quantity $\sum_{n=1}^N \varphi^4_{\nu n}$ is recognized as the
inverse participation ratio,~\cite{Thouless74} which is inversely
proportional to the localization size of the exciton state
$|\nu\rangle$. Thus, we see that pure dephasing is suppressed for
more extended states, an effect known as the motional
narrowing.~\cite{Wubs98,Malyshev98}

Like in the one-phonon assisted process, the scattering rate
between different states $|\nu\rangle$ and $|\mu\rangle$ is
proportional to the overlap of their site occupations. The final
expressions for the rates $W^{(2)}_{\mu\nu}$ resulting from
Eqs.~(\ref{1W2}) and (\ref{F2model}) are derived in Appendix A;
they depend on the sign of $\omega_{\mu\nu}$ as well as on the
relation between $|\omega_{\mu\nu}|$ and $\omega_c$. Distinction
is made between three inelastic channels: downward
($\downarrow\downarrow$), in which two phonons are emitted, cross
($\uparrow\downarrow$), in which one phonon is absorbed and another
is emitted, and upward ($\uparrow\uparrow$), in which two phonons
are absorbed. The fourth type of scattering is the elastic (pure
dephasing) channel, discussed above already. When calculating the
two-phonon assisted dephasing rate $\Gamma^{(2)}_\nu$ of state
$|\nu\rangle$, we will account for elastic as well as inelastic
contributions:
\begin{equation}
\label{Gamma2}
    \Gamma^{(2)}_\nu
    = \frac{1}{2}\Big[ W^{(2)}_{\nu\nu}
    + \sum_{\mu (\ne \nu)}W^{(2)}_{\mu\nu} \Big] \ .
\end{equation}
This rate depends on temperature as a result of the mean
occupation numbers $\bar{n}(\omega)$ and
$\bar{n}(\omega^{\prime})$ of the vibrational modes.

\section{Disorder-free aggregate}
\label{homogeneouslimit}

In order to gain insight in the temperature dependence of the
dephasing rates and the absorption bandwidth, it is useful to start
by considering a homogeneous aggregate, i.e., $\sigma=0$.
Analytical results can then be obtained if we restrict the resonant
interactions $J_{nm}$ to nearest-neighbor ones. In this
approximation (which we will relax in our numerical analysis), we
have
\begin{subequations}
\label{homogeneous}
\begin{equation}
\label{phimu}
    \varphi_{\nu n} = \left(\frac{2}{N+1}\right)^{1/2}
    \sin \frac{\pi\nu n}{N+1} \ ,
\end{equation}
\begin{equation}
\label{Emu}
    E_\nu = -2J \cos \frac{\pi \nu}{N+1} \ .
\end{equation}
\end{subequations}
The corresponding overlap integrals occurring in Eqs.~(\ref{W1})
and (\ref{W2}) now read
\begin{equation}
\label{overlap}
    \sum_{n=1}^N \varphi^2_{\mu n} \varphi_{\nu n}^2
    = \frac{1}{N+1}\left[ 1 + \frac{1}{2}\left(\delta_{\mu\nu}
    + \delta_{\mu+\nu,N+1}\right)\right] \ .
\end{equation}

\subsection{One-phonon-assisted dephasing} \label{1}

In a homogeneous linear chain, the lowest exciton state, $|\nu=1
\rangle$, contains almost all oscillator strength, thus dominating
the absorption spectrum.\cite{Knapp84,Fidder90} Therefore, the
dephasing rate of this state, $\Gamma^{(1)}_1$, is of primary
interest. As follows from Eq.~(\ref{Gamma1}), it is determined by
the sum over scattering rates to the other exciton states ($\mu \ne
1$), all of which are higher in energy.

In order to evaluate $\Gamma^{(1)}_1$, we replace the summation in
Eq.~(\ref{Gamma1}) by an integration, $\big(\sum_\mu \to
[(N+1)/\pi]\int dK \big)$, which is allowed if $N \gg 1$ and $T \gg
E_2 - E_1$. We will also assume that $T \ll J$, which implies that
the relevant exciton levels are those near the lower exciton band
edge, where $E_\mu = -2J + JK^2$ with $K = \pi\mu/(N+1)$. Changing
the integration variable to $x=JK^2/T$, using $E_\mu - E_1 \approx
JK^2$ and replacing the lower integration limit $\pi/(N+1)$ by
zero, we obtain
\begin{equation}
\label{Gamma1homo}
    \Gamma^{(1)}_1 = \frac{W^{(1)}_0}{4\pi}
    \left(\frac{T}{J}\right)^{\alpha + \frac{1}{2}} \,
    \int_0^{\omega_c/T} dx \, \frac{x^{\alpha - \frac{1}{2}}}{e^x - 1} \ .
\end{equation}

For temperatures $T \ll \omega_c$, the upper integration limit may
be extended to infinity, and we arrive at
\begin{equation}
\label{Gamma1T<}
    \Gamma^{(1)}_1 =
    \frac{W^{(1)}_0}{4\pi}\Gamma\left(\alpha+\frac{1}{2}\right)
    \zeta\left(\alpha+\frac{1}{2}\right)
    \left(\frac{T}{J}\right)^{\alpha + \frac{1}{2}} \ ,
\end{equation}
where $\Gamma(z)$ and $\zeta(z)$ are the gamma-function and the
Rieman zeta-function, respectively. Thus, for $T \ll \omega_c$ the
one-phonon-assisted dephasing rate shows a power-law temperature
dependence. Note that for our model of acoustic phonons ($\alpha =
3$), $\Gamma^{(1)}_1$ increases quite steeply, namely as $T^{7/2}$.
From numerical evaluation of $\Gamma^{(1)}_1$ for a homogeneous
chain with all dipole-dipole interactions, we have found that the
exponent 7/2 is increased to 3.85, mainly as a consequence of
logarithmic corrections in the exciton dispersion near the lower
band edge.\cite{Malyshev95,Didraga04} If we go beyond the parabolic
range of the energy spectrum, the growth becomes even steeper; the
exponent then tends to 4, because the density of states becomes a
constant towards the center of the band.

In the opposite limit $T \gg \omega_c$, the exponential in the
denominator of Eq.~(\ref{Gamma1homo}) can be expanded in a Taylor
series. Up to second order, one obtains
\begin{equation}
\label{Gamma1T>}
    \Gamma^{(1)}_1 =
    \frac{W^{(1)}_0}{2\pi (2\alpha-1)}
    \left(\frac{\omega_c}{J}\right)^{\alpha - \frac{1}{2}}\,\,
    \frac{T}{J} \ ,
\end{equation}
which simply reflects the linear high-temperature dependence of
the mean occupation number $\bar{n}(\omega)$. Obviously, this
scaling also holds in the presence of disorder.

\subsection{Pure dephasing} \label{2}

We now turn to the temperature dependence of the pure dephasing
rate $\Gamma^{(2)}_\nu = (1/2)W^{(2)}_{\nu\nu}$. As for a
homogeneous aggregate this rate does not depend on the state index
$\nu$, we will simply denote it as $\Gamma^{(2)}_\mathrm{pure}$.
Using the explicit form of ${\cal F}^{(2)}(\omega,\omega^{\prime})$
given by Eq.~(\ref{Wnunu}), we arrive at
\begin{equation}
\label{gp}
    \Gamma^{(2)}_\mathrm{pure} = \frac{3}{2}\frac{W^{(2)}_0}{N+1}
    \left(\frac{T}{J}\right)^{2\alpha +1} \int_0^{\omega_c/T}
    \mathrm{d}x \, \frac{x^{2\alpha}\, e^x}{(e^x - 1)^2}  \ .
\end{equation}
From Eq.~(\ref{gp}) it follows that for $T \ll \omega_c$ \,
($\omega_c/T \to \infty$),
\begin{subequations}
\label{gp_limits}
\begin{equation}
\label{gpT<}
    \Gamma^{(2)}_\mathrm{pure} = \frac{3}{2}\frac{W^{(2)}_0}{N+1} \,
    \Gamma(2\alpha+1)\, \zeta(2\alpha) \left(\frac{T}{J}\right)^{2\alpha +1} \ ,
\end{equation}
while for $T \gg \omega_c$ \, ($\omega_c/T \to 0$),
\begin{equation}
\label{gpT>}
    \Gamma^{(2)}_\mathrm{pure} = \frac{3}{2}\frac{W^{(2)}_0}{(2\alpha-1)(N+1)}
    \left(\frac{\omega_c}{J}\right)^{2\alpha-1}
    \left(\frac{T}{J}\right)^{2}  \ .
\end{equation}
\end{subequations}

For the case of scattering on acoustic phonons ($\alpha = 3$) and
$T \ll \omega_c$, we thus arrive at $\Gamma^{(2)}_\mathrm{pure}
\propto T^7$. This temperature dependence resembles that for the
pure dephasing of an isolated state of a point center, derived by
McCumber and Sturge.~\cite{McCumber63} The only difference is that
the exciton dephasing rate undergoes suppression by a factor of
$N+1$ due to the motional narrowing effect. We note that this
narrowing is not observed for one-phonon-assisted dephasing [see
Eqs.~(\ref{Gamma1T<}) and~(\ref{Gamma1T>})]. The $T^2$-scaling of
$\Gamma^{(2)}_\mathrm{pure}$ in the high-temperature limit ($T \gg
\omega_c$) results from the square of the mean phonon occupation
number involved in Eq.~(\ref{Wnunu}).

To conclude this section, we stress that the $T^{2\alpha+1}$ and
$T^2$ scaling relations of the pure dephasing rate with
temperature obtained here, also hold for disordered aggregates,
because this result is determined only by the two-vibration
spectral density ${\cal F}^{(2)}(\omega,\omega^{\prime})$. The
suppression factor, however, will then be determined by the
exciton localization size, rather than the chain length.

\subsection{Inelastic two-phonon-assisted dephasing} \label{3}

Finally, we analyze the dephasing rate of the superradiant state
($\nu=1$) resulting from the two-phonon inelastic scattering of
excitons. This rate, which will be denoted as
$\Gamma^{(2)}_\mathrm{inel}$, is determined by the sum of
scattering rates to all higher states, $\Gamma^{(2)}_\mathrm{inel}
= (1/2)\sum_{\mu \ne 1} W^{(2)}_{\mu 1}$. Using Eqs.~(\ref{1W2})
and (\ref{F2model}), and making the same assumptions as in the case
of one-phonon-assisted dephasing (Sec.~\ref{1}), we obtain
\begin{eqnarray}
\label{Gamma2_inel}
    \Gamma^{(2)}_\mathrm{inel} & = & \frac{W^{(2)}_0}{4\pi}
    \left( \frac{T}{J} \right)^{2\alpha + \frac{3}{2}}
    \int_0^{\omega_c/T}\mathrm{d}x
    \int_0^{\omega_c/T}\mathrm{d}y
\nonumber\\
\nonumber\\
    & \times & \frac{x^\alpha\,y^\alpha}{(e^x - 1)(e^y - 1)}
\nonumber\\
\nonumber\\
    & \times &
    \left[ \,
    \frac{2e^y }{\sqrt{x - y}}\,\, \Theta(x - y)
     +  \frac{1}{\sqrt{x + y}} \,
    \right] \ .
\end{eqnarray}

In the low-temperature limit, $T \ll \omega_c$ \, ($\omega_c/T \to
\infty$), Eq.~(\ref{Gamma2_inel}) yields
\begin{equation}
\label{Gamma2_inelT<}
    \Gamma^{(2)}_\mathrm{inel} =  \frac{\kappa W^{(2)}_0}{4\pi}
    \left( \frac{T}{J} \right)^{2\alpha + \frac{3}{2}} \ ,
\end{equation}
where the numerical factor $\kappa$ is given by the double
integral in Eq.~(\ref{Gamma2_inel}), with both upper limits
replaced by infinity. Comparing Eq.~(\ref{Gamma2_inelT<}) with
Eq.~(\ref{gpT<}), we see that $\Gamma^{(2)}_\mathrm{inel}$ is
characterized by a steeper temperature dependence than
$\Gamma^{(2)}_\mathrm{pure}$. In particular, for $\alpha = 3$ the
rate $\Gamma^{(2)}_\mathrm{inel} \propto T^{15/2}$. This means
that at higher temperatures the inelastic two-phonon channel of
dephasing can compete with the elastic one (Sec.~\ref{2phonon}).
In the high-temperature limit, $T \gg \omega_c$ \, ($\omega_c/T
\to 0$), the rate $\Gamma^{(2)}_\mathrm{inel}$ is proportional to
$T^2$, which is the same scaling relation as in the case of pure
dephasing [Eq.~(\ref{gpT>})].

\section{Disordered aggregates}
\label{numerics}

In this section, we will analyze the temperature dependence of the
various dephasing contributions in the presence of disorder and
their effect on the width of the J-band. As we will see (and
already anticipated in Sec.~\ref{linear}), the dephasing rates are
spread over a wide region, in particular at low temperature. As a
consequence, it is not clear {\it a priori} what value (mean,
typical, or other) of these rates should be related to the
homogeneous width of the absorption spectrum. Therefore, the width
of the J-band was obtained by direct simulation of the absorption
spectrum, using the calculated dephasing rates to broaden each of
the exciton transitions in the band. Explicitly, we have
\begin{equation}
\label{A}
    A(E) = \frac{1}{N} \left\langle \sum_\nu  \frac{F_\nu}{\pi} \,\,
    \frac{\Gamma_\nu}{(E - E_\nu)^2 + \Gamma_\nu^2}\right\rangle \ .
\end{equation}
where $F_\nu = (\sum_{n=1}^N \varphi_{\nu n})^2$ is the
dimensionless oscillator strength of the $\nu$th exciton state and
$\Gamma_\nu = \gamma_\nu/2 + \Gamma^{(1)}_\nu + \Gamma^{(2)}_\nu$
is the total homogeneous width of this state. Here, $\Gamma^{(1)}$
and $\Gamma^{(2)}$ are given by Eqs.~(\ref{Gamma1})
and~(\ref{Gamma2}), respectively, and $\gamma_\nu = \gamma_0
F_\nu$ is the radiative decay rate of state $\nu$ ($\gamma_0$
denotes the radiative constant of a monomer). In all the
simulations, we will assume the limit $\omega_c \gg T$, which
implies that $\omega_c \to \infty$ in Eqs.~(\ref{F1model})
and~(\ref{F2model}). As before, the angular brackets denote the
average over the random realizations of the site energies
$\{\varepsilon_n\}$. The resulting J-bandwidth $\Delta$ was
determined as the full width at half maximum of the thus
calculated absorption spectrum. A similar approach has been
used~\cite{Basko03} to simulate the absorption spectrum of THIATS
aggregates~\cite{Scheblykin96} at room temperature.

\subsection{One-phonon-assisted dephasing}
\label{1phonon}

\begin{figure}[ht]
\includegraphics[width= \columnwidth,clip]{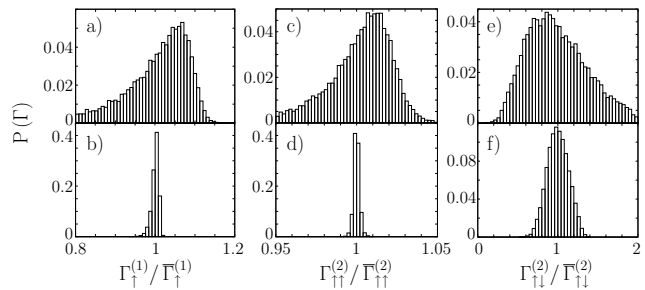}
\caption{Normalized distributions of the one-phonon-assisted and
two-phonon-assisted dephasing rates $\Gamma^{(1)}_{\uparrow}$ [(a)
and (b)], $\Gamma^{(2)}_{\uparrow\uparrow}$ [(c) and (d)], and
$\Gamma^{(2)}_{\uparrow\downarrow}$ [(e) and (f)] of the lowest
exciton state in each one of $3 \times 10^4$ disorder realizations
for chains of 500 molecules. Upper panels correspond to $T = 0.06
J$, while lower panels were calculated for $T = 0.23 J$. All plots
were obtained for a disorder strength $\sigma = 0.135 J$ and a
one- or two-vibration spectral density given by
Eq.~(\ref{F1model}) or Eq.~(\ref{F2model}) with $\alpha=3$.  Note
that the horizontal axes are scaled by the average values of the
three rates considered, indicated as
$\overline{\Gamma}^{(1)}_{\uparrow}$,
$\overline{\Gamma}^{(2)}_{\uparrow\uparrow}$, and
$\overline{\Gamma}^{(2)}_{\uparrow\downarrow}$. These averages
depend on temperature.} \label{g1}
\end{figure}

\begin{figure}[ht]
\includegraphics[width= \columnwidth,clip]{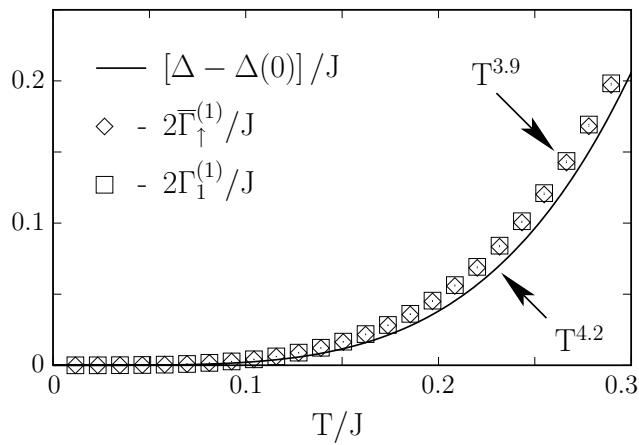}
\caption{Temperature dependence of two measures for the
homogeneous width for disordered aggregates ($\sigma = 0.1 J$)
compared to the dephasing rate of the lowest (superradiant)
exciton state in a homogeneous aggregate ($\sigma=0$; squares).
For the disordered case, we plotted the dynamic part $\Delta(T) -
\Delta(0)$ of the total J-bandwidth (solid line) and the average
value $\overline{\Gamma}^{(1)}_{\uparrow}$ of the dephasing rate
of the lowest exciton state found in each disorder realization
(diamonds). In all cases we used a chain length of $N=500$, a
monomer radiative rate $\gamma_0 = 1.5 \times 10^{-5} J$, and a
one-phonon spectral density of the form Eq.~(\ref{F1model}) with
$W_0^{(1)} = 25 J$ and $\alpha=3$.} \label{g1_fwhm}
\end{figure}

In order to study fluctuations in the dephasing rates, we analyzed
their statistics, focusing on the lowest exciton state for each
randomly generated disorder realization. This choice was motivated
by the fact that the low-lying states dominate the absorption
spectrum. The dephasing rate of the lowest state is denoted
$\Gamma^{(1)}_{\uparrow}$; its distribution, collected by
considering $3\times 10^4$ disorder realizations for chains of
$N=500$ molecules and a disorder strength $\sigma = 0.135 J$ is
presented in Figs.~\ref{g1}(a) and~\ref{g1}(b). In generating
these figures, we used a one-vibration spectral density of the
form Eq.~(\ref{F1model}), with $\alpha=3$. Furthermore, the rates
were calculated for $T = 0.06 J$ [Fig.~\ref{g1}(a)] and $T = 0.23
J$ [Fig.~\ref{g1}(b)]; for the protoptypical aggregates of
pseudoisocyanine ($J=600$ cm$^{-1}$), this agrees with
temperatures of 50 K and 200 K, respectively. We note that the
horizontal axis of the distributions is scaled by the average
value $\overline{\Gamma}^{(1)}_{\uparrow}$, so that the value of
$W_0^{(1)}$ does not affect the figures. Of course, this scaling
renders the axis temperature dependent, because
$\overline{\Gamma}^{(1)}_{\uparrow}$ strongly depends on $T$, as
we will see below (Fig.~\ref{g1_fwhm}).

These figures clearly demonstrate that the relative spread in
$\Gamma^{(1)}_{\uparrow}$ may be considerable, in particular at
low temperatures. This may be understood from the local band-edge
level structure of the disordered tight-binding
Hamiltonian.~\cite{Malyshev91} For temperatures smaller than the
J-bandwidth, the exciton scatters between discrete levels in the
vicinity of the band edge that are localized in the same region of
the chain. Both the energy spacing between these states and their
localization size undergo large fluctuations: $\delta E_{\mu\nu}
\sim E_{\mu\nu}$ and $\delta (\sum_{n=1}^N \varphi_{\mu n}^2
\varphi_{\nu n}^2) \sim \sum_{n=1}^N \varphi_{\mu n}^2
\varphi_{\nu n}^2$.~\cite{Malyshev01} As a consequence, $\delta
W^{(1)}_{\mu\nu} \sim W^{(1)}_{\mu\nu}$. When the temperature is
increased, the exciton in the lowest state may scatter to many
higher-lying states, which often are delocalized over an
appreciable part of the chain. This smears the fluctuations that
occur in the scattering rates between individual states, leading
to a decrease in the relative spread of the dephasing rate of the
lowest state.

\begin{figure}[ht]
\includegraphics[width= \columnwidth,clip]{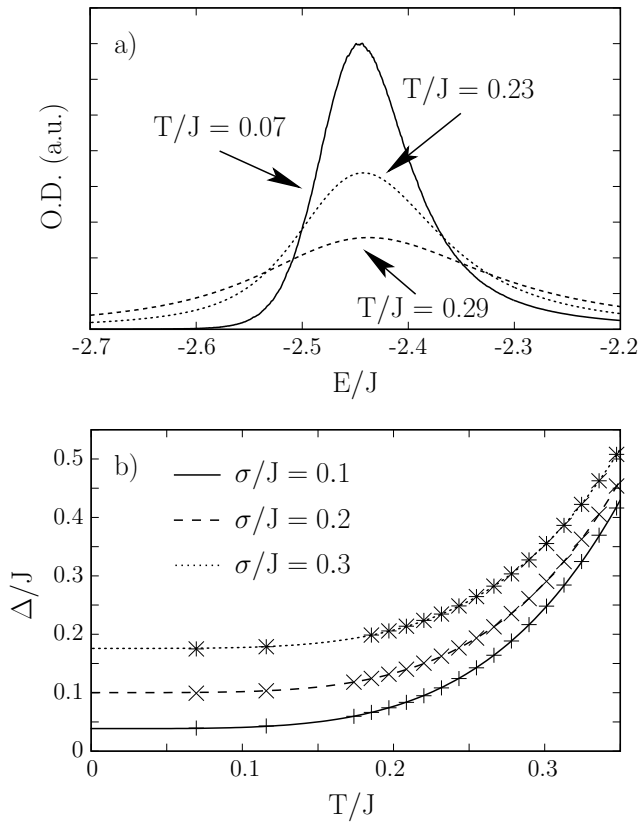}
\caption{(a) Calculated J-band for $T =0.07 J$, $0.23 J$, and
$0.29 J$, at a fixed disorder strength $\sigma = 0.2 J$. A
one-phonon spectral density was used of the form
Eq.~(\ref{F1model}), with $W_0^{(1)} = 25 J$ and $\alpha = 3$,
while the monomer radiative rate was set to $\gamma_0 = 1.5 \times
10^{-5} J$. (b) Temperature dependence of the width $\Delta(T)$ of
the calculated J-band (symbols) for three values of the disorder
strength, $\sigma = 0.1 J$, $0.2 J$, and $0.3 J$. Other parameters
as in (a). The solid, dashed and dotted curves represent the
corresponding fits, according to Eq.~(\ref{fit}).}
\label{p3nocutoff}
\end{figure}

We next turn to the absorption band calculated according to
Eq.~(\ref{A}), neglecting the role of two-phonon scattering
[$W_0^{(2)}=0$]. In Fig.~\ref{p3nocutoff}(a) this band is plotted
for three temperatures at a fixed disorder strength of $\sigma
=0.2 J$. For the one-phonon spectral density we used the form
Eq.~(\ref{F1model}) with $W_0^{(1)}=25 J$ and $\alpha = 3$, and
took $\gamma_0 = 1.5 \times 10^{-5} J$, which is typical for
J-aggregates of polymethine dyes. Chains of $N = 500$ molecules
were considered. The simulated spectra clearly demonstrate the
thermal broadening, caused by growing homogeneous widths of the
individual exciton transitions. At low temperature, the
homogeneous broadening is negligible, the J-band is inhomogeneous,
with a width that is determined by the disorder strength. With
growing temperature, the J-band becomes more homogeneous, as is
apparent from the fact that it gets more symmetric.

In Fig.~\ref{p3nocutoff}(b) we plotted by symbols the simulated
J-band width $\Delta(T)$ as a function of temperature for three
values of the disorder strength: $\sigma = 0.1 J$, $0.2 J$, and
$0.3 J$ [all other parameters were taken as in
Fig.~\ref{p3nocutoff}(a)]. As is seen, the $\Delta(T)$ shows a
plateau at the value of the inhomogeneous width, $\Delta(0) = 0.04
J$, $0.1 J$, and $0.18 J$, respectively. Beyond these plateaus,
$\Delta(T)$ goes up quite steeply, reflecting the fact that the
homogeneous (dynamic) broadening becomes dominant. To accurately
extract at low temperatures the small homogeneous contribution to
the total width, we generated up to $4 \times 10^5$ disorder
realizations. At higher temperatures, this number could be
restricted to 4000, owing to the reduction of the relative
fluctuations (cf.~Fig.~\ref{g1}).

Inspired by the analytically obtained power-laws for the
one-phonon-assisted dephasing rate as a function of temperature
for homogeneous aggregates (Sec.~\ref{1}), we considered a
parametrization of the form
\begin{equation}
\label{fit}
    \Delta(T) = \Delta(0) + a W^{(1)}_0 \left(T/J\right)^{p}
\end{equation}
for the total band width in disordered aggregates. It turned out
that the calculated line widths as a function of temperature could
be fitted very well by the relation~(\ref{fit}) [curves in
Fig.~\ref{p3nocutoff}(b)]. The corresponding fit parameters are $a
= 1.24$ and $p = 4.16$ for $\sigma = 0.1 J$, $a = 1.32$ and $p =
4.29$ for $\sigma = 0.2 J$, and $a = 1.20$ and $p = 4.27$ for
$\sigma = 0.3 J$. The scaling relation Eq.~(\ref{fit}) turns out
to hold over an even wider range of $\sigma$ and $W_0$
values.\cite{Heijs05} This implies that, although the J-band is
built up from a distribution of exciton states with different
dephasing rates, the total width $\Delta(T)$ may effectively be
separated in an inhomogeneous width, $\Delta(0)$, and a dynamic
contribution.

We note that the fitting exponent $p$ is larger than the value
3.85 found in the absence of disorder (Sec.~\ref{1}). This
increase results from downward scattering processes between
optically dominant exiton states, which are possible in the
presence of disorder, but not for the superradiant state in the
homogeneous chain. This claim may be substantiated by considering
the dephasing rates of the lowest exciton state of each disorder
realization. We numerically generated the average of this
quantity, $\overline{\Gamma}^{(1)}_{\uparrow}$, from $3 \times
10^4$ disorder realizations for chains of $N=500$ molecules with
$\sigma =0.1 J$, $W_0^{(1)} =25 J$, and $\alpha=3$. This average
is shown as a function of temperature in Fig.~\ref{g1_fwhm}
(diamonds), together with the dynamic contribution to the total
J-bandwidth, $\Delta(T) - \Delta(0)$ (solid line), and the
dephasing rate $\Gamma^{(1)}_{1}$ of the superradiant state for a
homogeneous chain of the same length (squares). As is seen, the
dynamic part $\Delta(T) - \Delta(0)$ has a larger exponent $p =
4.16$ than $\overline{\Gamma}^{(1)}_{\uparrow}$ $(p = 3.85)$. It
is remarkable, however, that $\overline{\Gamma}^{(1)}_{\uparrow}$
and $\Gamma^{(1)}_{1}$ display almost identical behavior, at least
in the relevant region $T \gtrsim \Delta(0)$, where the
homogeneous contribution to the J-bandwidth is noticeable. For $T
\ll \Delta(0)$, where $\Gamma^{(1)}_{\uparrow}$ undergoes
significant fluctuations (see Fig.~\ref{g1}),
$\overline{\Gamma}^{(1)}_{\uparrow}$ turns out to be much smaller
than $\Gamma^{(1)}_{1}$. In the high-temperature regime, these
fluctuations are washed out and the two quantities become almost
identical.

\begin{figure}[ht]
\includegraphics[width= \columnwidth,clip]{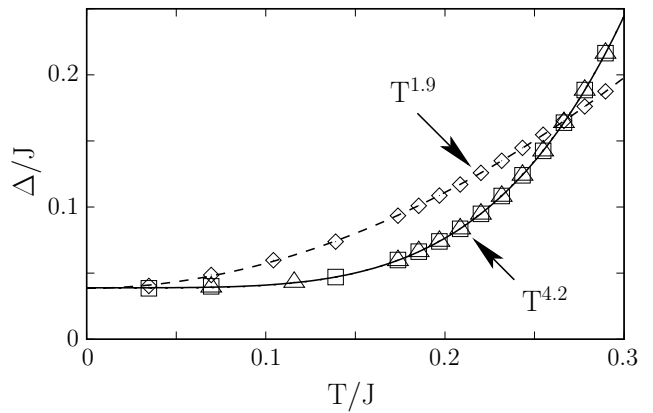}
\caption{Temperature dependence of
the calculated width $\Delta(T)$ using three different models for
the one-phonon spectral density. Diamonds (triangles) correspond
to a spectral density of the form Eq.~(\ref{F1model}) with
$\alpha=1$ and $W_0^{(1)} = 5 J$ ($\alpha=3$ and $W_0^{(1)} = 25
J$), while the squares are the results for ${\cal F}^{(1)}(\omega)
= W_0^{(1)}(\omega/J)^3 [1 + \sin(2\pi\omega/\tilde{\omega})]$
with $W_0^{(1)} = 25 J$ and $\tilde{\omega} = J/6$. The various
curves are fits to the power-law Eq.~(\ref{fit}). In all cases we
used a chain length of $N=500$ and a monomer radiative rate
$\gamma_0 = 1.5 \times 10^{-5} J$.} \label{width_sin}
\end{figure}

So far, we have only presented numerical results for a one-phonon
spectral density Eq.~(\ref{F1model}) with the power $\alpha=3$,
which corresponds to a Debye model for the host vibrations. To end
this subsection, we will address the effect of changing the
spectral density. First, we consider the effect of the value for
$\alpha$. The diamonds in Fig.~\ref{width_sin} present our
results for the temperature dependence of the width $\Delta(T)$
obtained for $\alpha=1$ and $W_0 = 5 J$. As before, we found that
these data may be fitted by a simple power-law of the form
Eq.~(\ref{fit}) (dashed line). In this case we find $p=1.9$,
which, again, is slightly larger than the value 3/2 found from
Eq.~(\ref{Gamma1T<}), due to the correction of the dispersion
relation arising from the long-range dipole-dipole interactions
and downward scattering processes that contribute to the total
J-bandwidth. Clearly, these data demonstrate the sensitivity of
the temperature dependence of the total linewidth to the power
$\alpha$.

Interestingly, it turns out that while the temperature dependence
of the J-bandwidth is sensitive to the overall frequency scaling
of the spectral density, it is not sensitive to fluctuations of
this scaling around an average power-law. To demonstrate this, we
have considered a spectral density of the form ${\cal
F}^{(1)}(\omega) = W_0^{(1)}(\omega/J)^3 [1 +
\sin(2\pi\omega/\tilde{\omega})]$, which only on average exhibits
an $\omega^3$-dependence. The results for the width $\Delta(T)$
obtained for $W_0^{(1)} = 25 J$ and $\tilde{\omega} = J/6$ are
presented as squares in Fig.~\ref{width_sin}. Remarkably, these
results are indistinguishable from those obtained without
fluctuations (i.e., $\tilde{\omega} = \infty$: triangles). The
reason is that at elevated temperatures the function $\omega^3
\bar{n}(\omega)$ varies slowly on the scale of $\tilde{\omega}$,
so that the modulating function $1 +
\sin(2\pi\omega/\tilde{\omega})$ may be replaced by its average
value, which equals unity.

\subsection{Two-phonon-assisted dephasing}
\label{2phonon}

\begin{figure}[ht]
\includegraphics[width= \columnwidth,clip]{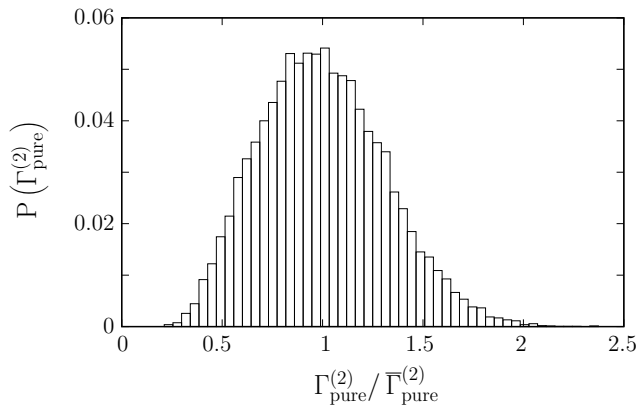}
\caption{As in Fig.~\ref{g1}, but now for
$\Gamma^{(2)}_\mathrm{pure}$. As is discussed in the text, this
distribution is solely due to fluctuations in the inverse
participation ratio of the exciton states and does not depend on
temperature.} \label{g2p}
\end{figure}

\begin{figure}[ht]
\includegraphics[width= \columnwidth,clip]{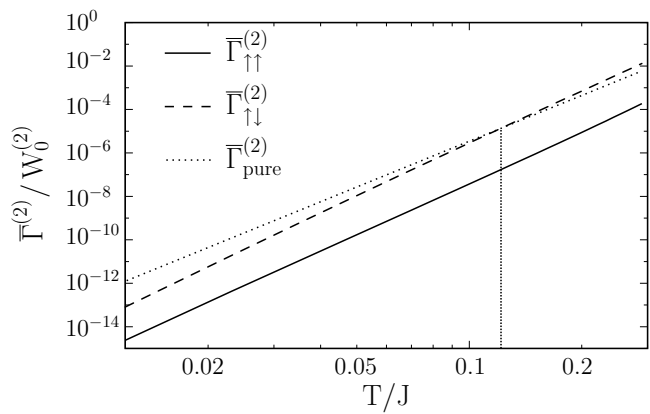}
\caption{Temperature dependence of the average values
($\overline{\Gamma}^{(2)}_{\uparrow\uparrow}$,
$\overline{\Gamma}^{(2)}_{\uparrow\downarrow}$, and
$\overline{\Gamma}^{(2)}_\mathrm{pure}$) of the three
contributions to the two-phonon-assisted dephasing rate of the
lowest exciton state in disordered aggregates of $N=500$
molecules. Parameters were chosen as in Fig.~\ref{g1}. The
vertical dotted line shows the crossing point at $T = 0.12 J$.}
\label{gamma_tmp}
\end{figure}

As we have seen in Sec.\ref{quadratic}, the two-phonon-assisted
dephasing rate $\Gamma_\nu^{(2)}$ consists of four
contributions, one of which is elastic (indicated as ``pure", as
it is responsible for pure dephasing), while the other three are
inelastic and are indicated as downward ($\downarrow\downarrow$),
cross ($\uparrow\downarrow$), and upward ($\uparrow\uparrow$). In
Figs.~\ref{g1}(c)-~\ref{g1}(f), \ref{g2p}, and \ref{gamma_tmp} we
present results for the statistics of these various contributions
for disordered aggregates. In all cases, we used chains of 500
molecules, a disorder strength of $\sigma = 0.135 J$, and a
two-vibration spectral density of the form Eq.~(\ref{F2model}),
with $\alpha=3$. The statistics are presented for the lowest
exciton state in each one of $3 \times 10^4$ randomly generated
disorder realizations. For this state the downward contribution
vanishes and the two-phonon-assisted dephasing rate reads
$\Gamma_\nu^{(2)} = \Gamma^{(2)}_{\uparrow\uparrow} +
\Gamma^{(2)}_{\uparrow\downarrow} + \Gamma^{(2)}_\mathrm{pure}$.
These three remaining contributions were calculated using the
expressions derived in the Appendix.

From Figs.~\ref{g1}(c) and~\ref{g1}(d) we observe that at a given
temperature, the relative spread in
$\Gamma^{(2)}_{\uparrow\uparrow}$ is much smaller than that in
$\Gamma^{(2)}_{\uparrow\downarrow}$. The reason is that in a
two-phonon-assisted upward process the exciton in the lowest state
scatters to more higher-energy states than in a cross process. As a
result, fluctuations in $\Gamma^{(2)}_{\uparrow\uparrow}$ are
suppressed more than those in $\Gamma^{(2)}_{\uparrow\downarrow}$.
Upon heating, the spread of both $\Gamma^{(2)}_{\uparrow\uparrow}$
and $\Gamma^{(2)}_{\uparrow\downarrow}$ reduces, which has the
same explanation as given for this effect in the case of
one-phonon-assisted dephasing (Sec.~\ref{1phonon}). The
distribution of $\Gamma^{(2)}_\mathrm{pure}$ does not depend on
temperature at all, because, according to Eq.~(\ref{Wnunu}), the
rate $\Gamma^{(2)}_\mathrm{pure}$ fluctuates exclusively due to
fluctuations in the inverse participation ratio $\sum_{n=1}^N
\varphi_{\nu n}^4$. As the latter quantity is subject to large
fluctuations,~\cite{Malyshev01} this also explains the large
relative spread in $\Gamma^{(2)}_\mathrm{pure}$.

In Fig.~\ref{gamma_tmp} we plotted (on a log-log scale) the
temperature dependence of the mean values
$\overline{\Gamma}^{(2)}_{\uparrow\uparrow}$,
$\overline{\Gamma}^{(2)}_{\uparrow\downarrow}$, and
$\overline{\Gamma}^{(2)}_\mathrm{pure}$, obtained from averaging
over these rates for the lowest exciton states in the simulations
discussed above. This figure nicely shows the relative importance
of the different dephasing channels. We clearly see that
$\overline{\Gamma}^{(2)}_{\uparrow\uparrow} \ll
\overline{\Gamma}^{(2)}_{\uparrow\downarrow},
\overline{\Gamma}^{(2)}_\mathrm{pure}$, i.e., the inelastic
channel of dephasing due to double phonon absorption is
inefficient, at all temperatures. More importantly, we observe
that the cross channel of inelastic two-phonon-assited dephasing
successfully competes with the pure dephasing contribution. At low
temperatures, pure dephasing dominates, while at higher
temperatures the inelastic cross process is more important. This
correlates well with our findings for disorder-free aggregates
[see discussion below Eq.~(\ref{Gamma2_inelT<})]. For the disorder
strength considered here, the cross-over occurs at $T_0 = 0.12 J$
($\approx 100$ K for $J = 600$ cm$^{-1}$). We note that when
considering two-phonon scattering, one often restricts to
modelling the elastic process (see, e.g.,
Ref.~\onlinecite{Kuhn97}). From the above we see that this is not
justified at elevated temperatures.

Despite the fact that not all three curves in Fig.~\ref{gamma_tmp}
are exactly straight lines (deviations occur in the unimportant
low-temperature part), they all can be fitted very well by a
power-law, $a(T/J)^p$. In doing so, we obtained
\begin{subequations}
\begin{equation}
\label{barGmm}
    \overline{\Gamma}^{(2)}_{\uparrow\uparrow} = 0.18 W_0^{(2)}
    \left(T/J\right)^{8.3} \ ,
\end{equation}
\begin{equation}
\label{barGpm}
    \overline{\Gamma}^{(2)}_{\uparrow\downarrow} = 9.01 W_0^{(2)}
    \left(T/J\right)^{7.9} \ ,
\end{equation}
\begin{equation}
\label{barGpure}
    \overline{\Gamma}^{(2)}_\mathrm{pure} = 1.88 W_0^{(2)}
    \left(T/J\right)^{7} \ .
\end{equation}
\end{subequations}
We recall that the exponent $p = 7$ in the last formula is an
exact result for $\alpha= 3$ and $\omega_c \gg T$, as was argued
at the end of Sec.~\ref{2} already. As in the case of one-phonon
scattering, we see that the inelastic two-phonon dephasing rates
exhibit a steeper temperature dependence than for the homogeneous
aggregate with nearest-neighbor interactions
[Eq.~(\ref{Gamma2_inelT<})].

\section{Comparison to experiment and discussion} \label{experiment}

It is of interest to see to what extent the model we presented
here is able to explain the temperature dependence of the
homogeneous broadening in molecular aggregates. As mentioned in
the Introduction, Renge and Wild~\cite{Renge97} found that the
total J-bandwidth $\Delta(T)$ of PIC-Cl and PIC-F over a wide
temperature range (from 10 K to 300 K) follows a power-law scaling
as in Eq.~(\ref{fit}). Although the power reported by these
authors ($p=3.4$) is smaller than the ones we derived in
Sec.~\ref{1phonon}, from direct comparison to the experimental
data we have found that a model of one-phonon scattering with a
spectral density given by Eq.(\ref{F1model}) with $\alpha = 3$ and
$\omega_c \rightarrow \infty$, yields an excellent quantitative
explanation of the experiments over the entire temperature range,
both for the shape and the width of the J-band. The same turns out
to be true for the J-bandwidth of PIC-Br, measured between 1.5 K
and 180 K.~\cite{Fidder90} In all these fits, $\sigma$ and $W_0$
were the only two free parameters that could be adjusted to
optimize comparison to experiment. Details will be published
elsewhere,\cite{Heijs05} together with a fit of the much
debated\cite{Fidder90,deBoer89,Spano90,Fidder95,Potma98}
temperature dependence of the fluorescence lifetime of these
aggregates.

\begin{figure}[ht]
\includegraphics[width= \columnwidth,clip]{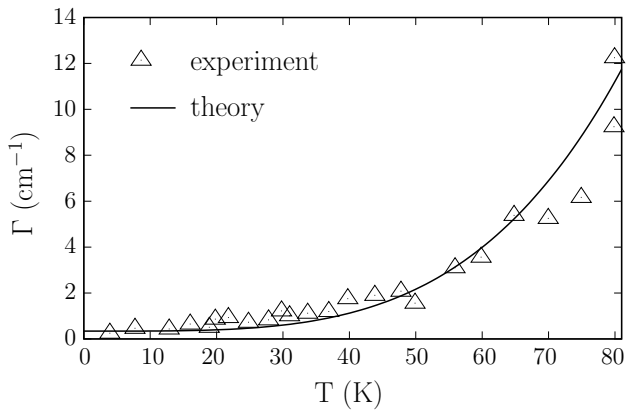}
\caption{Hole width $\Gamma$ as a function of temperature measured
for aggregates of PIC-I after pumping in the center of the J-band
(triangles).\cite{Hirschmann89} The solid line is our fit,
assuming one-quantum scattering on acoustic phonons in the host.
Model parameters are discussed in the text.
} \label{holeburning}
\end{figure}

Here, we present an explicit comparison to the hole-burning data
reported by Hirschmann and Friedrich.\cite{Hirschmann89} Using
this technique, they measured the homogeneous width of the exciton
states in the center of the J-band for PIC-I over the temperature
range 350 mK to 80 K. Their data for the holewidth $\Gamma$ are
reproduced as triangles in Fig.~\ref{holeburning}. The solid line
shows our fit to these data, obtained by simulating disordered
chains of $N=250$ molecules with a one-phonon spectral density of
the form Eq.~(\ref{F1model}) with $\alpha=3$ and $\omega_c
\rightarrow \infty$. The resonant interaction strength and the
monomer radiative rate were chosen at the accepted values of
$J=600$ cm$^{-1}$ and $\gamma_0 = 1.5 \times 10^{-5}J = 2.7 \times
10^8$ s$^{-1}$, respectively. Thus, the only free parameters were
the disorder strength $\sigma$ and scattering strength $W_0^{(1)}$.
First, we fixed the value of $\sigma$ by fitting the
low-temperature (4 K) absorption spectrum, where the homogeneous
broadening may be neglected. This yielded $\sigma = 0.21 J$. Next,
$W_0^{(1)}$ was adjusted such that the measured growth of the hole
width was reproduced in an optimal way. Thus, we found
$W_0^{(1)}=180J$. For each given temperature the simulated hole
width was obtained as the average of the dephasing rate of the
excitons found in an interval of 0.06 cm$^{-1}$ around the center
of the simulated J-band at that temperature.

Taking into account the fair amount of scatter in the experimental
data, we conclude from Fig.~\ref{holeburning} that our model
yields a good fit to the measurements. As mentioned above, the
same holds for the J-bandwidth in aggregates of PIC-Cl, PIC-F, and
PIC-Br. This yields valuable information about the dominant
mechanism of dephasing in these materials. We conclude that this
mechanism is one-phonon-assisted scattering of excitons on
vibrations in the host characterized by a spectral density which
(on average) scales as $\omega^3$; this scaling yields a natural
explanation of the power-law thermal broadening of the J-band
found in various experiments. We stress that it is impossible to
fit the experimental data with a spectral density that is constant
or scales linearly with $\omega$, as that yields considerably
different power laws for the width [cf.~Fig.~\ref{width_sin}].
The $\omega^3$ scaling of the spectral density needed to fit the
experiments strongly suggests that acoustic phonons dominate the
scattering process. Thus, the spectral width measured over a broad
temperature range is an excellent probe for the scattering
mechanism.

It is appropriate to comment on the value of $W_0^{(1)}$ obtained
from our fit, which seems to be very large. It should be kept in
mind that $W_0^{(1)}$ is a phenomenological scattering strength,
which combines several microscopic material properties [see
discussion below Eq.~(\ref{F1model})]. Most importantly, the value
found here is consistent with a perturbative treatment of the
scattering process: the scattering rates between the optically
dominant states obtained from it turned out to be much smaller
than their energy separation.

We finally address an alternative mechanism of dephasing, namely
scattering on local vibrations belonging to the aggregate. This has
been suggested by several authors based on activation-law fits of
the measured homogeneous contribution to the
J-bandwidth.~\cite{Fidder90,Hirschmann89} To get an estimate
whether this is a reasonable mechanism, let us neglect the
disorder and make the nearest-neighbor approximation for the
resonance interactions $J_{nm}$. Then, $E_\nu$ and $\varphi_{\nu
n}$ are given by Eq.~(\ref{homogeneous}). Close to the lower
exciton band edge, the region of our interest, $E_\nu = -2J +
J\pi^2\nu^2/(N+1)^2$. Furthermore, let us parameterize the spectral
function of a local vibration of frequency $\omega_0$ as ${\cal
F}(\omega) = 2\pi V_0^2 \delta(\omega - \omega_0)$, where $V_0$ is
the coupling constant to the excitons. We are interested in the
dephasing rate $\Gamma_1$ of the superradiant state $|\nu=1
\rangle$. Using the above simpifications and replacing in
Eq.~(\ref{Gamma1}) the summation over exciton states by an
integration, one easily arrives at
\begin{equation}
\label{arrenius}
     \Gamma_1 = V_0^2 \frac{\bar n(\omega_0)}{(J \omega_0)^{1/2}}.
\end{equation}
Comparing this result to the activation law $b\exp(-\omega_0/T)$,
used in Refs.~\onlinecite{Fidder90} and~\onlinecite{Hirschmann89}
to fit the experimental data, one obtains $V_0^2 =
b(J\omega_0)^{1/2}$. Substituting the values $b = 3000$ cm$^{-1}$
and $\omega_0 = 330$ cm$^{-1}$ from
Ref.~\onlinecite{Hirschmann89}, and $J = 600$ cm$^{-1}$, we obtain
as estimate for the exciton-phonon coupling $V_0 \approx 1100$
cm$^{-1}$. This is an enormously large number. In particular, the
Stokes losses $S = V_0^2/\omega_0 \approx 3700$ cm$^{-1}$ turn out
to be much larger than the resonant interaction $J = 600$
cm$^{-1}$. Under these conditions, a strong exciton self-trapping
is to be expected, resulting in a reduction of the exciton
bandwidth by a factor of $\exp(-S/\omega_0) \ll
1$.~\cite{Rashba82} Coherent motion of the exciton, even over a
few molecules, is then hardly possible: any disorder will destroy
it. This observation is not consistent with the widely accepted
excitonic nature of the aggregate excited states, as corroborated
by many optical and transport
measurements.\cite{Kobayashi96,Knoester02}

\section{Concluding remarks}
\label{concl}

In this paper we presented a theoretical study of the temperature
dependence of the exciton dephasing rate in linear J-aggregates
and the resulting width of the total absorption band (the J-band).
As dephasing mechanism we considered scattering of the excitons on
vibrations of the host matrix, taking into account both one- and
two-vibration scattering. The excitons were obtained from
numerical diagonalization of a Frenkel exciton Hamiltonian with
energy disorder and their dephasing rates were subsequently
calculated using a perturbative treatment of the exciton-vibration
interaction (Fermi Golden Rule).

In the absence of disorder, the lowest (superradiant) exciton
state dominates the absorption spectrum. As a result, the dephasing
rate of this state directly gives the homogeneous width of the
J-band. We analytically calculated the temperature dependence of
this homogeneous width for both one- and two-vibration scattering,
assuming a Debye-like model for the host vibration density of
states. It turned out that in all cases the homogeneous width
obeys a power-law as a function of temperature, with the value of
the exponent depending on the shape of the low-energy part of the
vibronic spectrum (Sec.~\ref{homogeneouslimit}).

In the presence of disorder the optically dominant exciton states
still reside close to the bottom of the band, but their energies
are now spread and their wave functions become localized on finite
segments of the chain.  We have found that the various one- and
two-phonon-induced contributions to the dephasing rate undergo
significant fluctuations, because the exciton energies and overlap
integrals vary considerably from one disorder realization to the
other (Sec.~\ref{numerics}). As a consequence, one cannot use the
dephasing rates of individual states to characterize the
homogeneous broadening of the J-band. Instead, we  simulated the
total J-band and showed that its width effectively separates in an
inhomogeneous (zero-temperature) contribution and a dynamic
(homogeneous) part. The latter scales with temperature according
to a power law [Eq.~(\ref{fit})], with an exponent that is
somewhat larger than the one found for the homogeneous broadening
in the absence of disorder. We also showed that amongst the
two-vibration scattering processes, inelastic channels will at
elevated temperatures dominate the usually considered
pure-dephasing contribution.

Finally, from comparison to absorption and hole-burning
experiments (Sec.~\ref{experiment}), we found that the dominant
mechanism of dephasing for J-aggregates lies in one-phonon
scattering of excitons on vibrations of the host matrix,
characterized by a spectral density which (on average) scales like
the third power of the phonon frequency. This suggests that
acoustic phonons of the host play an important role in the
scattering process. All temperature dependent data available to
date, are consistent with this picture. By contrast, we argued
that the previously suggested mechanism of scattering on local
vibrations of the aggregate, leading to an activated thermal
behavior, is not consistent with the overwhelming amount of
evidence that the optical excitations in J-aggregates have an
excitonic character.

\appendix
\section{Two-phonon scattering rates}

In this Appendix we present expressions for the two-phonon-assisted
scattering rate $W^{(2)}_{\mu\nu}$, starting from Eq.~(\ref{1W2}).
After substituting the two-vibration spectral density given in
Eq.~(\ref{F2model}) and performing several algebraic
manipulations, we arrive at
\begin{eqnarray}
  \label{expression_W2}
    W^{(2)}_{\mu\nu}
    & = & W_0^{(2)} \left(\frac{T}{J}\right)^{2p+1}
    \sum_{n=1}^N \varphi^2_{\mu n} \varphi_{\nu n}^2
\nonumber\\
\nonumber\\
    & \times & \Big[F^{(2)}_{\downarrow\downarrow}(\omega_{\mu\nu})
    + 2F^{(2)}_{\uparrow\downarrow}(\omega_{\mu\nu})
    + F^{(2)}_{\uparrow\uparrow}(\omega_{\mu\nu})\Big] \ ,
\nonumber\\
\end{eqnarray}
where the temperature dependent functions
$F^{(2)}_{\downarrow\downarrow}(\omega_{\mu\nu})$,
$F^{(2)}_{\uparrow\downarrow}(\omega_{\mu\nu})$, and $
F^{(2)}_{\uparrow\uparrow}(\omega_{\mu\nu})$ distinguish between
the scattering processes in which two phonons are emitted
($\downarrow\downarrow$), one is absorbed and one is emitted
($\uparrow\downarrow$), and two phonons are absorbed
($\uparrow\uparrow$). They are given by
\begin{subequations}
\begin{eqnarray}
  \label{auxilary}
    F^{(2)}_{\downarrow\downarrow}(\omega_{\mu\nu})
    & = &
    \int_0^{\omega_c/T} dx
    \int_0^{\omega_c/T} dy
\nonumber\\
\nonumber\\
    & \times &
    f_{\downarrow\downarrow}(x,y) \,
    \delta \left(\frac{\omega_{\mu\nu}}{T} + x + y \right) \ ,
\\
\nonumber\\
    F^{(2)}_{\uparrow\downarrow}(\omega_{\mu\nu})
    & = &
    \int_0^{\omega_c/T} dx
    \int_0^{\omega_c/T} dy
\nonumber\\
\nonumber\\
    & \times &
    f_{\uparrow\downarrow}(x,y) \,
    \delta\left(\frac{\omega_{\mu\nu}}{T} - x + y \right) \ ,
\\
\nonumber\\
    F^{(2)}_{\uparrow\uparrow}(\omega_{\mu\nu})
    & = &
    \int_0^{\omega_c/T} dx
    \int_0^{\omega_c/T} dy
\nonumber\\
\nonumber\\
    & \times &
    f_{\uparrow\uparrow}(x,y) \,
    \delta\left(\frac{\omega_{\mu\nu}}{T} - x - y \right) \ ,
\end{eqnarray}
\end{subequations}
where, after changing to dimensionless integration variables $x =
\omega_q/T$ and $y = \omega_{q^{\prime}}/T$, we introduced the
auxiliary functions
\begin{subequations}
\begin{eqnarray}
    f_{\downarrow\downarrow}(x,y)
    & = &
    \frac{x^p e^x}{e^x-1} \,
    \frac{y^p e^y}{e^y - 1} \ ,
\\
\nonumber\\
    f_{\uparrow\downarrow}(x,y)
    & = &
    \frac{x^p}{e^x - 1} \,
    \frac{y^p e^y}{e^y - 1} \ ,
\\
\nonumber\\
    f_{\uparrow\uparrow}(x,y)
    & = &
    \frac{x^p}{e^x-1} \,
    \frac{y^p}{e^y - 1} \ .
\end{eqnarray}
\end{subequations}
We note that for $\mu = \nu$ ($\omega_{\mu\nu} = 0$), the only non
vanishing term is $F^{(2)}_{\uparrow\downarrow}(0)$, which
describes the elastic channel of scattering (pure dephasing).

We now further analyze the three contributions to the scattering
rate. As ${\cal F}^{(2)}_{\downarrow\downarrow}(\omega_{\mu\nu})$
describes the emission of two vibrational quanta, we have
$\omega_{\mu\nu} < 0$. Thus, performing the $y$ integration, we
obtain
\begin{subequations}
\begin{eqnarray}
    F^{(2)}_{\downarrow\downarrow}(\omega_{\mu\nu})
    =
    \int_0^{-\omega_{\mu\nu}/T}{\rm d}x \,
    f_{\downarrow\downarrow}\left(x, -\frac{\omega_{\mu\nu}}{T} - x\right) \ ,
\end{eqnarray}
if $-\omega_{\mu\nu} < \omega_c$, and
\begin{eqnarray}
    F^{(2)}_{\downarrow\downarrow}(\omega_{\mu\nu})
    =
    \int_{-(\omega_{\mu\nu} + \omega_c)/T}^{\omega_c/T}{\rm d}x \,
    f_{\downarrow\downarrow}\left(x, -\frac{\omega_{\mu\nu}}{T} - x\right) \ ,
\nonumber
\\
\end{eqnarray}
\end{subequations}
if $\omega_c < -\omega_{\mu\nu} < 2\omega_c$, while
$F^{(2)}_{\downarrow\downarrow}(\omega_{\mu\nu})=0$ otherwise.

Next, the contribution which involves the emission and absorption
of one vibrational quantum, is given by
\begin{subequations}
\begin{eqnarray}
    F^{(2)}_{\uparrow\downarrow}(\omega_{\mu\nu})
    =
    \int_{\omega_{\mu\nu}/T}^{\omega_c/T}{\rm d}x \,
    f_{\uparrow\downarrow}\left(x, -\frac{\omega_{\mu\nu}}{T} + x\right) \ ,
\end{eqnarray}
if $0 \leq \omega_{\mu\nu} < \omega_c$.  If $0 < -\omega_{\mu\nu}
< \omega_c$,
\begin{eqnarray}
    F^{(2)}_{\uparrow\downarrow}(\omega_{\mu\nu})
    =
    \int_0^{(\omega_c+\omega_{\mu\nu})/T}{\rm d}x \,
    f_{\uparrow\downarrow}\left(x, -\frac{\omega_{\mu\nu}}{T} + x\right) \ ,
\end{eqnarray}
\end{subequations}
and $F^{(2)}_{\uparrow\downarrow}(\omega_{\mu\nu})=0$ otherwise.

Finally, the contribution that results from the absorption of two
vibrational quanta ($\omega_{\mu\nu} > 0$), yields
\begin{subequations}
\begin{eqnarray}
    F^{(2)}_{\uparrow\uparrow}(\omega_{\mu\nu})
    =
    \int_0^{\omega_{\mu\nu}/T}{\rm d}x \,
    f_{\uparrow\uparrow}\left(x,\frac{\omega_{\mu\nu}}{T} - x\right) \ ,
\end{eqnarray}
if $0 < \omega_{\mu\nu} < \omega_c$, and
\begin{eqnarray}
    F^{(2)}_{\uparrow\uparrow}(\omega_{\mu\nu})
    =
    \int_{(\omega_{\mu\nu}-\omega_c)/T}^{\omega_c/T}{\rm d}x \,
    f_{\uparrow\uparrow}\left(x,\frac{\omega_{\mu\nu}}{T} - x\right) \ ,
\end{eqnarray}
if $\omega_c < \omega_{\mu\nu} < 2\omega_c$, while
$F^{(2)}_{\uparrow\uparrow}(\omega_{\mu\nu})=0$ otherwise.
\end{subequations}

\end{document}